\documentclass[12pt]{article}%
\usepackage{amsmath,latexsym}
\usepackage{graphicx}
\usepackage{amsmath}
\usepackage{amsfonts}
\usepackage{amssymb}%
\begin{document}

\title{{Wormholes supported by small extra dimensions}}
   \author{
Peter K. F. Kuhfittig*\\  \footnote{kuhfitti@msoe.edu}
 \small Department of Mathematics, Milwaukee School of
Engineering,\\
\small Milwaukee, Wisconsin 53202-3109, USA}

\date{}
 \maketitle

\begin{abstract}\noindent
Holding a Morris-Thorne wormhole open requires 
a violation of the null energy condition, calling 
for the need for so-called exotic matter near the 
throat.  Many researchers consider exotic matter 
to be completely unphysical in classical general 
relativity.  It has been shown, however, that 
the existence of an extra macroscopic dimension 
can resolve this issue: the throat could be lined 
with ordinary matter, while the extra dimension 
is then responsible for the unavoidable energy 
violation.  The purpose of this paper is to show 
that the extra dimension can be microscopic, a 
result that is consistent with string theory.
   
\end{abstract}

\section{Introduction}\label{E:introduction}
This paper is concerned with a number of 
fundamental issues in the study of Morris-Thorne 
wormholes, even raising the question whether 
a basic wormhole structure can even be 
hypothesized.  Moreover, while they may be 
just as good a prediction of Einstein's theory 
as black holes, wormholes are subject to severe 
restrictions from quantum field theory, in 
particular, the need to violate the null energy 
condition, calling for the existence of ``exotic 
matter" to hold a wormhole open.  It has been 
shown, however, that this requirement can be 
met via the existence of an extra macroscopic 
spatial dimension.  It is proposed in this 
paper that the extra dimension can be extremely 
small, an appoach that is consistent with string 
theory.  Furthermore, the existence of the extra 
dimension would allow the throat of the wormhole 
to be lined with ordinary matter, while the 
unavoidable violation of the null energy condition 
can be attributed to the higher spatial dimension. 

The other issue to be addressed is the enormous 
radial tension at the throat of any 
moderately-sized wormhole, a problem that is 
usually ignored.

\section{Wormhole structure}\label{S:structure} 
Wormholes have been a subject of interest 
ever since it was realized that the 
Schwarzschild solution and therefore black 
holes can be viewed as wormholes, albeit 
nontraversable.  More recently, the problem 
of entanglement has drawn attention to a 
special type of wormhole, the Einstein-Rosen 
bridge, to explain this phenomenon.  
Accordingly, we will assume that a basic 
wormhole structure can be hypothesized.

While there had been some forerunners, what 
came to be called a Morris-Thorne wormhole 
was first proposed by Morris and Thorne 
\cite{MT88}, who proposed the following static 
and spherically symmetric line element for a 
wormhole spacetime:
\begin{equation}\label{E:line1}
ds^{2}=-e^{2\Phi(r)}dt^{2}+e^{2\lambda(r)}dr^2
+r^{2}(d\theta^{2}+\text{sin}^{2}\theta\,
d\phi^{2}),
\end{equation}
where
\begin{equation}
   e^{2\lambda(r)}=\frac{1}{1-\frac{b(r)}{r}}.
\end{equation}
(We are using units in which $c=G=1$.)  The 
terminology introduced in Ref. \cite{MT88} 
has become standard: $\Phi=\Phi(r)$ is called
the \emph{redshift function}; this function
must be finite everywhere to prevent the 
occurrence of an event horizon. The function 
$b=b(r)$ is called the \emph{shape function} 
since it determines the spatial shape of the 
wormhole whenever it is depicted in an embedding 
diagram \cite{MT88}.  The spherical surface 
$r=r_0$ is called the \emph{throat} of the 
wormhole.  In a Morris-Thorne wormhole, the
shape function must satisfy the following
conditions: $b(r_0)=r_0$, $b(r)<r$ for $r>r_0$,
and $b'(r_0)<1$, called the \emph{flare-out
condition} in Ref. \cite{MT88}.  In classical
general relativity, the flare-out condition can
only be met by violating the null energy condition
(NEC), which states that for the energy-momentum
tensor $T_{\alpha\beta}$
\begin{equation}
   T_{\alpha\beta}k^{\alpha}k^{\beta}\ge 0\,\,
   \text{for all null vectors}\,\, k^{\alpha}.
\end{equation}
Matter that violates the NEC is called ``exotic"
in Ref. \cite{MT88}; the term is borrowed from 
quantum mechanics.  To see the effect on 
wormholes, consider the radial outgoing null
vector $(1,1,0,0)$, which yields
\begin{equation}
   T_{\alpha\beta}k^{\alpha}k^{\beta}=\rho+
      p_r<0
\end{equation}
whenever the NEC is violated.  Here 
$T^t_{\phantom{tt}t}=-\rho(r)$
is the energy density, $T^r_{\phantom{rr}r}=
p_r(r)$ is the radial pressure, and
$T^\theta_{\phantom{\theta\theta}\theta}=
T^\phi_{\phantom{\phi\phi}\phi}=p_t(r)$
is the lateral (transverse) pressure.  
Another requirement is \emph{asymptotic
flatness:}
\begin{equation}
   \text{lim}_{r\rightarrow\infty}\Phi(r)=0
   \quad \text{and} \quad
   \text{lim}_{r\rightarrow\infty}
   \frac{b(r)}{r}=0.
\end{equation}

For later reference, let us now state the 
Einstein field equations in the orthonormal 
frame:
\begin{equation}\label{E:Einstein}
   G_{\hat{\alpha}\hat{\beta}}=R_{\hat{\alpha}\hat{\beta}}-\frac{1}
{2}Rg_{\hat{\alpha}\hat{\beta}}=8\pi T_{\hat{\alpha}\hat{\beta}},
\end{equation}
where
\begin{equation}
   g_{\hat{\alpha}\hat{\beta}}=
   \left(
   \begin{matrix}
   -1&0&0&0\\
   \phantom{-}0&1&0&0\\
   \phantom{-}0&0&1&0\\
   \phantom{-}0&0&0&1
   \end{matrix}
   \right).
\end{equation}
In the orthonormal frame, we can simply 
write $T_{00}=\rho$ and $T_{11}=p_r$.

\section{The exotic-matter problem}
As noted in the Introduction, holding a 
wormhole open requires a violation of 
the NEC, thereby calling for the existence 
of exotic matter, at least in the vicinity 
of the throat.  The problematical nature of 
exotic matter in classical general 
relativity has led to a certain skepticism: 
many researchers consider such wormhole 
solutions to be completely unphysical, 
thereby ruling out the existence of 
macroscopic traversable wormholes in 
Einstein's theory.  This has suggested 
solutions beyond the classical theory.  
For example, it was proposed by Lobo 
and Oliveira \cite{LO09} that in $f(R)$ 
modified gravity, the wormhole throat 
could be lined with ordinary matter, 
while the violation of the NEC can be 
attributed to the higher-order curvature 
terms.  Another possibility is to invoke 
noncommutative geometry, an offshoot of
string theory \cite{pK23, pK20}.

In this paper, we are going to be more 
interested in the effects of an extra 
spatial dimension, an example of which 
is the \emph{induced-matter theory} by 
P.S. Wesson \cite{WP92, pW13, pW15}:
what we perceive as matter is merely 
the impingement of the higher-dimensional 
space onto ours.  The relationship between 
the energy violation and the existence 
of extra dimensions is taken up in 
Ref. \cite{pK18}, where the extra 
dimensions are assumed to be macroscopic.
String theory, on the other hand, assumes 
the existence of extra small dimensions; 
these are sometimes referred to as 
``compactified" or ``curled up."  So 
it makes sense for us to follow suit 
and assume the existence of at least 
one small extra dimension.

\section{A small extra dimension}
If we are going to assume the existence 
of an extra dimension, we must first 
decide on the line element.  With 
Eq. (\ref{E:line1}) in mind, let us 
consider the form
\begin{equation}\label{E:line2}
ds^{2}=-e^{2\Phi(r)}dt^{2}+e^{2\lambda(r)}dr^2
+r^{2}(d\theta^{2}+\text{sin}^{2}\theta\,
d\phi^{2})+e^{2\mu(r,l)}dl^2,
\end{equation}
where $l$ is the extra coordinate.  This 
choice was motivated in part by symmetry 
considerations: all the exponential terms 
have the same form.

Our next step is to list the components 
of the Ricci tensor from Ref. \cite{pK18} 
in the orthonormal frame:
\begin{multline}\label{E:R1}
  R_{00}=-\frac{1}{2}\frac{d\Phi(r)}{dr}\frac{rb'-b}{r^2}
  +\frac{d^2\Phi(r)}{dr^2}\left(1-\frac{b}{r}\right)\\+
  \left[\frac{d\Phi(r)}{dr}\right]^2\left(1-\frac{b}{r}\right)
  +\frac{2}{r}\frac{d\Phi(r)}{dr}\left(1-\frac{b}{r}\right)
    +\frac{d\Phi(r)}{dr}\frac{\partial\mu(r,l)}{\partial r}
  \left(1-\frac{b}{r}\right),
\end{multline}
\begin{multline}\label{E:R2}
   R_{11}=\frac{1}{2}\frac{d\Phi(r)}{dr}\frac{rb'-b}{r^2}
  -\frac{d^2\Phi(r)}{dr^2}\left(1-\frac{b}{r}\right)\\
  -\left[\frac{d\Phi(r)}{dr}\right]^2\left(1-\frac{b}{r}\right)
  +\frac{rb'-b}{r^3}-\frac{\partial^2\mu(r,l)}{\partial r^2}
  \left(1-\frac{b}{r}\right)\\
  +\frac{1}{2}\frac{\partial\mu(r,l)}{\partial r}
  \frac{rb'-b}{r^2}-\left[\frac{\partial\mu(r,l)}{\partial r}
  \right]^2\left(1-\frac{b}{r}\right),
\end{multline}
\begin{equation}
   R_{22}=R_{33}=-\frac{1}{r}\frac{d\Phi(r)}{dr}
   \left(1-\frac{b}{r}\right)+\frac{1}{2}\frac{rb'-b}{r^3}
   +\frac{b}{r^3}
   -\frac{1}{r}\frac{\partial\mu(r,l)}{\partial r}
   \left(1-\frac{b}{r}\right),
\end{equation}
and
\begin{multline}\label{E:R3}
  R_{44}=-\frac{d\Phi(r)}{dr}\frac{\partial\mu(r,l)}{\partial r}
  \left(1-\frac{b}{r}\right)-\frac{\partial^2\mu(r,l)}
  {\partial r^2}\left(1-\frac{b}{r}\right)\\
  +\frac{1}{2}\frac{\partial\mu(r,l)}{\partial r}\frac{rb'-b}{r^2}
  -\left[\frac{\partial\mu(r,l)}{\partial r}\right]^2
  \left(1-\frac{b}{r}\right)
  -\frac{2}{r}\frac{\partial\mu(r,l)}{\partial r}
  \left(1-\frac{b}{r}\right).
\end{multline}

The Ricci tensor plays a key role in 
analyzing the wormhole solution.  Upon 
closer examination, it turns out that 
the function $\mu(r,l)$ never occurs as 
an exponent or as a factor, but only as 
a derivative. So the magnitude of 
$\mu(r,l)$ does not have an effect: 
what matters is the rate of change of 
$\mu(r,l)$ with respect to $r$.  As far 
as the Ricci tensor is concerned, 
$\mu(r,l)$ can have any magnitude and 
either algebraic sign.

Returning to line element (\ref{E:line2}), 
it now follows that if $\mu(r,l)$ is 
negative and large in absolute value, 
then $e^{\mu(r,l)}$ can be extremely 
small or even compactified (in the 
language of string theory).  So our basic 
assumption, the existence of an extra 
small spatial dimension, is consistent 
with string theory.  The reason is that 
the small size is the only property from 
string theory that we are making use of. 
This is going to lead to our main 
conclusion.

\section{The large radial tension}
   \label{S:large}
Before continuing, let us to return to Ref.
\cite{MT88} to consider another problem, 
the radial tension at the throat.  First 
we need to recall that the radial tension 
$\tau(r)$ is the negative of the radial 
pressure $p_r(r)$.  According to Ref. 
\cite{MT88}, the Einstein field equations 
can be rearranged to yield $\tau(r)$.  
Temporarily reintroducing $c$ and $G$, 
we obtain
\begin{equation}
   \tau(r)=\frac{b(r)/r-2[r-b(r)]\Phi'(r)}
   {8\pi Gc^{-4}r^2}.
\end{equation}
The radial tension at the throat therefore 
becomes
\begin{equation}\label{E:tau}
  \tau(r_0)=\frac{1}{8\pi Gc^{-4}r_0^2}\approx
   5\times 10^{41}\frac{\text{dyn}}{\text{cm}^2}
   \left(\frac{10\,\text{m}}{r_0}\right)^2.
\end{equation}
As noted in Ref. \cite{MT88}, for a throat size 
of $r_0=3$ km, $\tau(r)$ has the same magnitude
as the pressure at the center of a massive
neutron star.  This is enough to suggest that 
moderately-sized wormholes are actually compact 
stellar objects \cite{pK22}.  According to 
Eq. (\ref{E:tau}), however, wormholes with 
low tidal forces could only exist on very 
large scales, i.e., such wormholes require 
very large throat sizes.

\section{The main result}
We know from Sec. \ref{S:structure} that 
the NEC states that for the energy-momentum 
tensor $T_{\alpha\beta}$, $T_{\alpha\beta}
k^{\alpha}k^{\beta}\ge 0\,\,$ for all null 
vectors $k^{\alpha}$.  We also recall that 
an ordinary Morris-Thorne wormhole [line 
element (\ref{E:line1})] can only  be 
maintained if the NEC is violated.  In 
particular, for the outgoing null vector 
$(1,1,0,0)$, the violation reads 
\begin{equation}
   T_{\alpha\beta}k^{\alpha}k^{\beta}
   =\rho +p_r<0.
\end{equation}
Returning to Eq. (\ref{E:Einstein}), 
observe that 
\begin{equation}
   8\pi(\rho+p_r)=8\pi(T_{00}+T_{11})=
   \left[R_{00}-\frac{1}{2}R(-1)\right]+
   \left[R_{11}-\frac{1}{2}R(1)\right]=
   R_{00}+R_{11}.
\end{equation}
Since we are primarily interested in 
the vicinity of the throat, we assume 
that $1-b(r_0)/r_0=0.$  So it follows 
immediately from Eqs. (\ref{E:R1}) 
and (\ref{E:R2}) that
\begin{equation}
  8\pi(\rho +p_r)|_{r=r_0}=
  \frac{rb'-b}{r^3}\left. +\frac{1}{2}
   \frac{\partial(\mu(r,l)}{\partial r}
   \frac{rb'-b}{r^2}\right |_{r=r_0}  
  =\frac{b'(r_0)-1}{r_0^2}
  \left[1+\frac{r_0}{2}\frac{\partial\mu(r_0,l)}{\partial r}
  \right].
\end{equation}
Recalling that $b'(r_0)<1,$ we obtain
\begin{equation}
   \rho +p_r>0 \quad \text{at} \quad r=r_0
\end{equation}
provided that
\begin{equation}\label{E:condition1}
   \frac{\partial\mu(r_0,l)}{\partial r}
   <-\frac{2}{r_0}.
\end{equation}
So, thanks to the extra dimension, 
the NEC is satisfied at the throat, 
which can therefore be lined with 
ordinary matter.  It is interesting 
to note that if $\mu(r,l)$ is independent 
of $r$, so that 
$\partial\mu(r,l)/\partial r=0$, we get 
\begin{equation}
    \rho+p_r|_{r=r_0}=\frac{1}{8\pi}
    \frac{b'(r_0)-1}{r_0^2}<0,
\end{equation}
the usual condition for a Morris-Thorne 
wormhole.

Condition (\ref{E:condition1}) also plays 
a key role in maintaining the wormhole.  
To show this, consider the null vector 
$(1,0,0,0,1).$  Assuming that the Einstein 
field equations hold in the five-dimensional 
spacetime, we now have 
\begin{equation}
   G_{00}+G_{44}=8\pi(T_{00}+T_{44})=
   \left[R_{00}-\frac{1}{2}Rg_{00}\right]
   +\left[R_{44}-\frac{1}{2}Rg_{44}\right]
   =R_{00}+R_{44}
\end{equation}  
and given that $1-b(r_0)/r_0=0,$ we get 
from Eqs. (\ref{E:R1}) and (\ref{E:R3})
that 
\begin{equation}\label{E:condition2}
   R_{00}+R_{44}=\frac{1}{2}
   \frac{rb'-b}{r^2}\left[
   -\frac{d\Phi(r)}{dr}
   +\frac{\partial\mu(r,l)}{\partial r}
   \right].
\end{equation}    
 Since $\partial\mu(r_0,l)/\partial r 
 <-2/r_0$, the second factor on the 
 right side of Eq. (\ref{E:condition2}) 
 is positive if 
\begin{equation}\label{E:condition3}
   \frac{d\Phi(r_0)}{dr}=-A<
   \frac{\partial\mu(r_0,l)}{\partial r}
   <-\frac{2}{r_0},
\end{equation}   
which is similar to Condition 
(\ref{E:condition1}).  It follows that 
$\rho+p_r|_{r=r_0}<0$.  We conclude that 
the NEC is satisfied at the throat in the 
four-dimensional spacetime but violated 
in the five-dimensional spacetime.

To finish the discussion, we need to return 
to Sec. \ref{S:large} and recall that the 
wormhole we are considering can only exist 
on a very large scale, i.e., with a very 
large throat radius $r=r_0$.  Inequality 
(\ref{E:condition1}) therefore implies 
that $\partial\mu(r_0,l)/\partial r$ 
is close to zero, and hence
\begin{equation}\label{E:condition4}
   \frac{\partial}{\partial r}e^{\mu(r_0,l)}
   =e^{\mu(r,l)}\left.\frac{\partial\mu(r,l)} 
   {\partial r}\right |_{r=r_0}
   <e^{\mu(r_0,l)}\left(-\frac{2}{r_0}\right)
\end{equation} 
is also close to zero.  So from the perspective 
of the four-dimensional spacetime, the extra 
dimension is not only nonincreasing, the small 
absolute value of the derivative in
(\ref{E:condition4}) actually implies that 
the extra dimension is essentially constant.  
In summary, a sufficiently large throat size 
ensures that the radial tension remains low 
and that the size of the extra dimension 
remains fixed.

\emph{Remark:} For the NEC to be met near 
the throat in the four-dimensional case, 
the shape function has to meet the condition 
$b'(r_0)>1/3.$  [See Ref. \cite{pK18} for details.]

\section{Conclusion}    
 It has been argued that wormholes are 
 just as good a prediction of Einstein's 
 theory as black holes, but they are 
 subject to severe restrictions from 
 quantum field theory.  In particular, 
 to hold a wormhole open requires a 
 violation of the NEC, which means that 
 the throat of the wormhole has to be 
 lined with ``exotic matter."  The 
 problematical nature of exotic matter 
 has led many researchers to conclude 
 that such wormhole solutions are 
 completely unphysical in classical 
 general relativity, thereby ruling out  
 the existence of macroscopic traversable 
 wormholes.  It has been shown, however, 
 that the existence of an extra 
 macroscopic spatial dimension can 
 account for the unavoidable violation 
 of the NEC, while allowing the throat 
 of the wormhole to be constructed from 
 ordinary matter.  The purpose of this 
 paper is to show that the extra 
 dimension can be microscopic.  This 
 result is consistent with string 
 theory which assumes that the extra 
 dimensions are ``compactified" or 
 ``curled up."  Since the small size 
 is the only property from string 
 theory that we are making use of, 
 our result suggests that string 
 theory is able to support 
 traversable wormholes.

 Regarding the redshift function 
 $\Phi=\Phi(r)$, in the original 
 Morris-Thorne wormhole, this function 
 can be freely assigned.  In our 
 situation, however, the need to violate 
 the NEC has led to Condition 
 (\ref{E:condition2}), implying that 
 $\Phi(r)$ and $\mu(r,l)$ have to meet 
 similar conditions, namely Inequalities 
 (\ref{E:condition1}) and 
 (\ref{E:condition3}), respectively.
 Here the throat radius $r=r_0$ is 
 very large, thereby implying that 
 traversable wormholes can only exist 
 on very large scales.


\begin{thebibliography}{20}
\bibitem{MT88}M.S. Morris and K.S. Thorne,
   Wormholes in spacetime and their use for interstellar
   travel: A tool for teaching general relativity. 
   Am. J. Phys., 56 (1988) 395-412.
\bibitem{LO09}F.S.N. Lobo and M.A. Oliveira,
   Wormhole geometries in$f(R)$ modified theories of gravity.
   Phys. Rev. D, 80 (2009) 104012.
\bibitem{pK23}P.K.F. Kuhfittig, Macroscopic 
   noncommutative-geometry wormholes as emergent 
   phenomena. Letters in High Energy Physics (LHEP), 
   2023 (2023) 399. 
\bibitem{pK20}P.K.F. Kuhfittig, Noncommutative-geometry 
   wormholes without exotic atter. Adv. Stud. Theor. Phys.,
   14 (2020) 219-225.    
\bibitem{WP92}P.S. Wesson and J. Ponce de Le\'{o}n,
   Kaluza-Klein equations, Einstein's equations, and 
   an effective energy-momentum tensor. J. Math. Phys., 
   33 (1992) 3883-3887.
\bibitem{pW13}P.S. Wesson, Astronomy and the fifth 
   dimension. arXiv: 1301.0033.
\bibitem{pW15}P.S. Wesson, The status of modern 
   five-dimensional gravity. Int. J. Mod. Phys. D, 
   24 (2015) 1530001.
\bibitem{pK18}P.K.F. Kuhfittig, Traversable wormholes 
   sustained by an extra spatial dimension. Phys. Rev. D, 
   98 (2018) 064041.   
\bibitem{pK22}P.K.F. Kuhfittig, A note on wormholes 
   as compact stellar objects. Fundamental J. Mod. 
   Phys., 17 (2022) 63-70.         

\end{thebibliography}
\end{document}